\documentclass[pra,aps,showpacs,groupedaddress,superscriptaddress,twocolumn, numerical]{revtex4-1}

\usepackage[utf8]{inputenc}
\usepackage{color}
\usepackage{bbm} 
\usepackage{nicefrac}
\usepackage{amsfonts,amsmath,amssymb,stmaryrd}
\usepackage{braket}

\usepackage[autostyle=true]{csquotes}

\usepackage{tabularx}
\usepackage{multirow} 
\usepackage{hhline}
\usepackage{graphicx}
\usepackage{subfigure}  %
\usepackage{bbm} 
\usepackage{hyperref}
\usepackage{mathrsfs}
\usepackage{verbatim}
\usepackage{centernot}
\usepackage{ulem}
\usepackage{array}
\usepackage{cancel}
\usepackage{ifthen}
\usepackage{bm}	%
\usepackage{todonotes}
\usepackage{siunitx}
\usepackage{float}
\presetkeys{todonotes}{size=\tiny}{}

\InputIfFileExists{gitinfo-latex.inc}{}{}

\usepackage[acronym,nomain]{glossaries}

\usepackage{units}
\usepackage{upgreek}

\definecolor{azure}{rgb}{.4,.7,.7}

\newcommand{\ti}[1]{_\text{#1}}
\newcommand{\rb}{\ti{Rb}}
\newcommand{\cs}{\ti{Cs}}
\newcommand{\kB}{k}

\newcommand{\mrb}{M}
\newcommand{\mcs}{m}
\newcommand{\vrb}{V}
\newcommand{\vcs}{v}
\newcommand{\densrb}{n}
\newcommand{\numrb}{N}
\newcommand{\tempcloud}{T}
\newcommand{\crosssec}{\sigma}
\newcommand{\collrate}{\Gamma}
\DeclareMathOperator\erf{Erf}

\makeatletter
\newsavebox\myboxA
\newsavebox\myboxB
\newlength\mylenA
\newcommand*\xoverline[2][0.75]{%
	\sbox{\myboxA}{$\m@th#2$}%
	\setbox\myboxB\null%
	\ht\myboxB=\ht\myboxA%
	\dp\myboxB=\dp\myboxA%
	\wd\myboxB=#1\wd\myboxA%
	\sbox\myboxB{$\m@th\overline{\copy\myboxB}$}%
	\setlength\mylenA{\the\wd\myboxA}%
	\addtolength\mylenA{-\the\wd\myboxB}%
	\ifdim\wd\myboxB<\wd\myboxA%
	\rlap{\hskip 0.5\mylenA\usebox\myboxB}{\usebox\myboxA}%
	\else
	\hskip -0.5\mylenA\rlap{\usebox\myboxA}{\hskip 0.5\mylenA\usebox\myboxB}%
	\fi}
\makeatother

\begin{document}
	\title{Observation of individual tracer atoms in an   ultracold dilute  gas}
	
	\author{Michael Hohmann}
	\affiliation{Department of Physics and Research Center OPTIMAS, University of Kaiserslautern, Germany}
	
	\author{Farina Kindermann}
	\affiliation{Department of Physics and Research Center OPTIMAS, University of Kaiserslautern, Germany}
	
	\author{Tobias Lausch}
	\affiliation{Department of Physics and Research Center OPTIMAS, University of Kaiserslautern, Germany}
	
	\author{Daniel Mayer}
	\affiliation{Department of Physics and Research Center OPTIMAS, University of Kaiserslautern, Germany}
	\affiliation{Graduate School Materials Science in Mainz, Gottlieb-Daimler-Strasse 47, 67663 Kaiserslautern, Germany}
	
	\author{Felix Schmidt}
	\affiliation{Department of Physics and Research Center OPTIMAS, University of Kaiserslautern, Germany}
	\affiliation{Graduate School Materials Science in Mainz, Gottlieb-Daimler-Strasse 47, 67663 Kaiserslautern, Germany}
	\author{Eric Lutz}
	\affiliation{Department of Physics, Friedrich-Alexander-Universität Erlangen-Nürnberg, 91058 Erlangen, Germany}
	
	\author{Artur Widera}
	\affiliation{Department of Physics and Research Center OPTIMAS, University of Kaiserslautern, Germany}
	\affiliation{Graduate School Materials Science in Mainz, Gottlieb-Daimler-Strasse 47, 67663 Kaiserslautern, Germany}

	\date{\today}
	
	\begin{abstract}
	Understanding the motion of a tracer particle in a rarefied gas is of fundamental and practical importance. We report the experimental investigation of individual  Cs atoms impinging on a dilute cloud of ultracold Rb atoms with variable density. We study the nonequilibrium relaxation of the initial nonthermal state and detect the effect of single collisions which has eluded observation so far.	We show that after few collisions, the measured spatial distribution  of the light tracer atoms is correctly described by a  generalized  Langevin equation with a velocity-dependent friction coefficient, over a large range of Knudsen numbers.
		\end{abstract}

	\maketitle
Diffusion  is an essential  and omnipresent transport phenomenon in nature. The motion of a tagged particle in a fluid is determined by its mass $m$ and by the density of the fluid via the Knudsen number $K_{n}$ \cite{bir94}. In the regime of large densities $(K_{n} \ll 1)$, the interparticle collision frequency is high and the fluid may be treated as a continuum medium. On the other hand, for low densities $(K_{n} \gg 1)$, individual collisions matter and the discrete nature of the fluid is apparent. The only closed equation applicable to all values of $K_{n}$ is the Boltzmann equation for the phase-space distribution of the particles \cite{bir94,cer00}. Solutions of this nonlinear kinetic equation have been obtained in the extreme situations of vanishing and infinite Knudsen numbers, and in the Brownian limit of a heavy tracer particle, $m/M \gg1 $, where $M$ is the mass of the fluid particles \cite{bir94,cer00}. However, despite its central importance for  the foundations of statistical physics  and the study of e.g. fluid flows in the upper atmosphere and aerosols dynamics \cite{cer00,mun89,sha16}, much less is known, both theoretically and experimentally, about the long-standing problem of light tagged particles diffusing in a dilute gas at intermediate $K_{n}$ \cite{mas83}. 

An alternative and highly successful description of a tagged particle is offered by the Langevin equation \cite{cof96,gil12}. In this stochastic approach, Newton's equation of motion for the particle is extended by  a friction force and a fluctuating force that account for the interaction with the surrounding fluid. The Langevin equation enables  simple evaluation of the macroscopic properties of the diffusing particle, without the need to compute complicated collision integrals as in the Boltzmann equation. It is valid in the Brownian limit of a  heavy tracer particle, where the friction coefficient is independent of the velocity of the particle \cite{maz70}. 
Recent experimental studies of Brownian motion have been reported in gases \cite{blu06,li10} and liquids \cite{hua11,khe14}. However, the usual Langevin equation  with constant friction does not hold for light tracer particles in dilute gases \cite{fer07,fer14}.

\begin{figure*}
	\begin{center}
		\includegraphics[width=164mm]{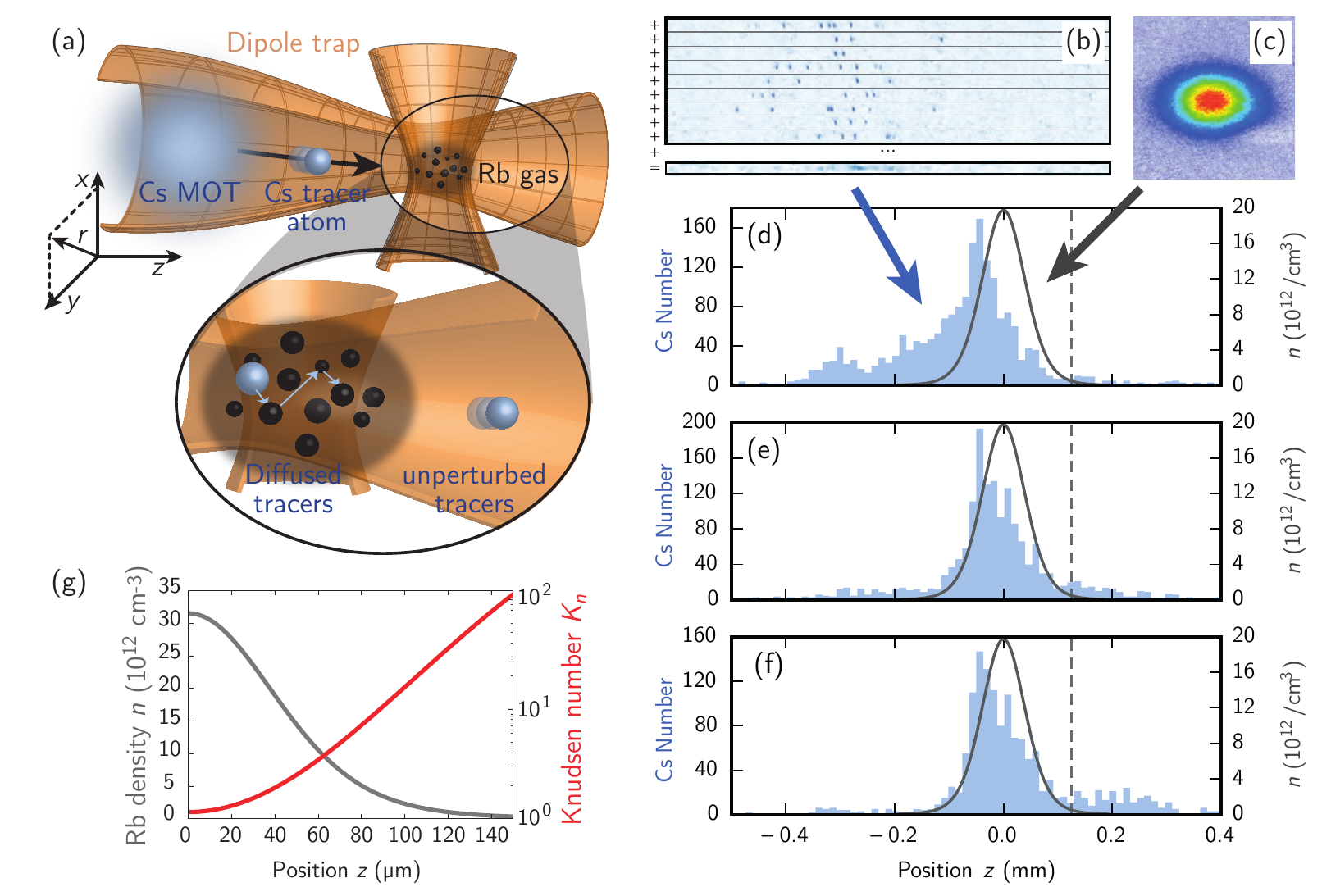}
	\end{center}
	\caption{\textbf{Sketch of experimental sequence}.
	(a) Individual Cs tracer atoms are released from a MOT at $z= -0.27$ mm and are drawn towards the center of a crossed dipole trap containing an ultracold thermal Rb cloud, where all atoms are in their respective hyperfine ground state. The Rb cloud contains typically $6\times10^{3}$  to   $3.5\times10^{4}$ atoms at a temperature of $\SI{2}{\micro\kelvin}$ and with ${1}/{e^2}$ widths of $\sigma_r=\SI{1.3}{\micro\meter}$ and $\sigma_z=\SI{39}{\micro\meter}$ in radial and axial directions.
	(b) After a variable (interaction) time $t$ after tracer release, the atomic motion is frozen by turning on a strong optical lattice. The position  of individual Cs atoms  is detected by fluorescence imaging after the Rb cloud has been removed.
	A series of fluorescence images for identical experimental conditions yields the spatial distribution of tracer atoms.
	(c) Time-of-flight image of the dilute Rb cloud.
	(d-f) Measured spatial tracer distribution for a central Rb density of $n=\SI{2e13}{\centi\meter^{-3}}$ for increasing interaction time $t$: (d) the Cs tracer distribution impinges onto the Rb cloud at \SI{17}{\milli\second}; (e) the Cs tracer distribution is axially overlapped with the Rb cloud at \SI{21}{\milli\second}; and (f) tracers which have not collided have left the region of the Rb cloud at \SI{25}{\milli\second}. 
	The grey Gaussian curve indicates the Rb cloud density, right vertical scale applies.
	The dashed vertical line at $z=0.125$ mm separates the diffused fraction that has thermalized inside the cloud (left) from the unperturbed fraction that has passed through the cloud (right).
	(g) Cloud density profile and corresponding Knudsen number variation for the highest central density of $n=3.2 \times 10^{13}\,$cm$^{-3}$ considered.
	}
	\label{fig:Sketch}
\end{figure*}

Here, we experimentally observe the motion of individual $^{133}$Cs atoms  impinging on a dilute cloud of ultracold $^{87}$Rb atoms as shown in Fig.~\ref{fig:Sketch}. We exploit the variable density of the cloud to explore a wide range of Knudsen numbers, from $K_{n}\simeq 1$ at the center to arbitrarily large values at the edges, for the light  Cs tracer atom with $m/M\simeq 1.5$. 
By initially  accelerating  the laser cooled Cs atoms to a nonthermal kinetic energy of about $43 k T$, we are  able to investigate in detail the nonequilibrium relaxation induced by the collisions with the cloud particles, from ballistic  to diffusive motion ($T$ is here the temperature of the thermal Rb cloud and $k$ the Boltzmann constant). The high resolution of our setup further allows to detect the effect of a single collision on the dynamics of the tracer atom. Discrete hard-sphere collision simulations reveal that only  few collisions suffice to thermalize a Cs atom to the cloud temperature. We additionally demonstrate that the measured spatial distribution of tagged  Cs atoms is well described, without any free parameters,  by a generalized Langevin equation with a velocity-dependent friction coefficient \cite{fer07,fer14}.  To our knowledge, this extension of the Langevin equation has never been experimentally verified so far.

In our experiment, laser cooled $^{133}$Cs atoms are released from a magneto-optical trap (MOT) into a crossed dipole trap containing a dilute  gas of  $^{87}$Rb atoms (Fig.~\ref{fig:Sketch}(a,b,c)) \cite{met02}.
After entering the cloud, a Cs atom undergoes collisions with the cold Rb atoms and thermalizes.
Ultracold temperatures in the micro Kelvin range lead to slow dynamics, both for the gas and the tracer atoms, with thermal velocities around $10\,$mm s$^{-1}$.  
While at room temperature, heavy tracer particles are bombarded at an extremely high rate (of the order of $10^{16}$ Hz in air \cite{maz02}), ultra low temperatures, a small mass ratio of the order of unity and low gas densities reduce this collision rate to values between $13 \times 10^3$ Hz at the center and $0$ at the edges of the gas cloud. Hence the effective mean free time between two collisions is in the experimentally accessible range of $\geq 0.1\,$ms. 
For the finite size cloud, there is a non-zero probability that a Cs atom does not collide at all and moves unperturbed through the cloud  (Fig.~\ref{fig:Sketch}(f)).
In order to observe the spatial distribution of the tagged Cs atoms, we freeze their positions after a given (interaction) time $t$ after release from the MOT by turning on a strong 1D optical lattice in the $z$-direction. We subsequently remove the Rb cloud from the trap and record the atomic position with fluorescence imaging (Fig.~\ref{fig:Sketch}(b)) \cite{Hohmann2015, Hohmann2016}. The spatial distribution for any fixed time is determined by accumulating atomic positions over about 600 realizations. The time evolution of the distribution is obtained by varying the interaction time $t$ (Figs.~\ref{fig:Sketch} (d)-(f)). We use only a few Cs atoms ($\approx7$ on average) per experimental run. Interspecies effects such as self-diffusion and self-thermalization are therefore negligible. Moreover, the thermal de Broglie wavelength  is smaller than the interparticle distance for the densities considered \cite{sup} and the dynamics of the atoms can hence be described classically.

Figures 2(a)-(c) show the measured Cs spatial distributions for three increasing values of the peak Rb density $n\ti{max}$ after an interaction  time $t=25$ ms. For low Rb density (Fig.~2(a)), we observe a bimodal distribution with a  fraction of unperturbed Cs atoms (right) that has passed through the Rb cloud (grey Gaussian curve) and a larger fraction  of diffused  atoms that has thermalized inside the cloud (left). The existence of the unperturbed fraction reveals the discrete nature of the dilute Rb gas. When the Rb density is increased (Figs.~2(b)-(c)), the unperturbed fraction shrinks, as the collision probability grows with $n\ti{max}$, and the diffused fraction does not fully penetrate the cloud. This may be qualitatively understood by noting that diffusion slows down with increasing density so that Cs tracer atoms need more time to reach the center of the cloud. In addition, three-body losses become important close to the center, where the density is highest, further decreasing the number of Cs atoms. A measurement of the three-body loss rate is presented in Ref.~\cite{sup}. The lifetime of a Cs atom at the peak of the spatial distribution in Fig.~2(c) is $\tau=3$ ms. Three-body-losses are therefore accounted for in all simulations.

\begin{figure}[t]
	\begin{center}
		\includegraphics{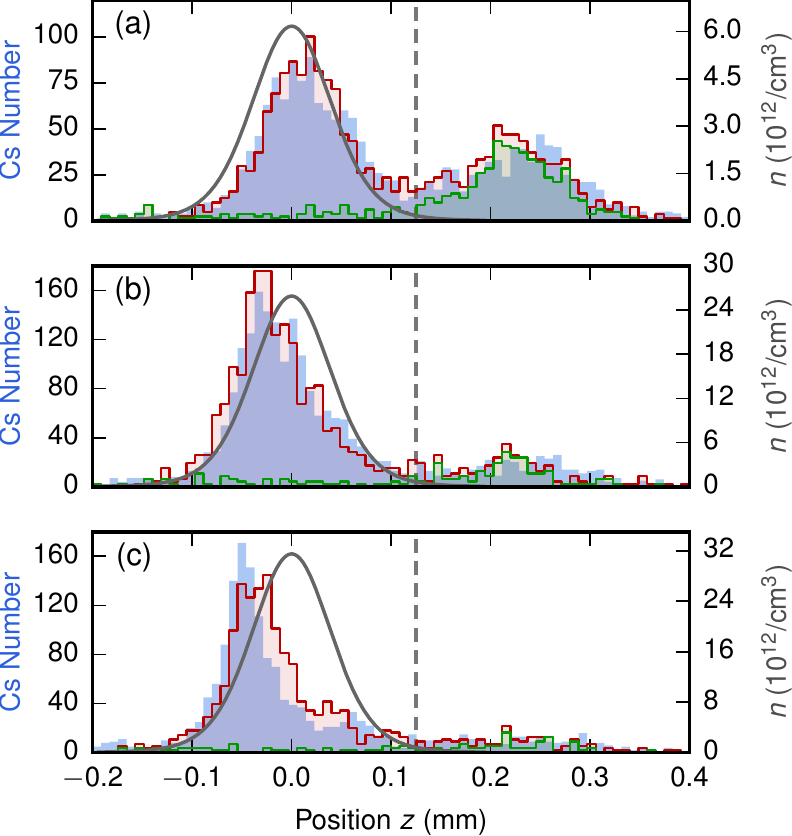}
	\end{center}
	\caption{\textbf{Spatial distribution of Cs tracer atoms}. Measured (blue shaded) and collision simulated (red line) spatial distributions after an interaction time of $25\,$ms for Rb center densities of (a) $n=6\times 10^{12}\,$cm$^{-3}$, (b) $n=2.6\times 10^{13}\,$cm$^{-3}$, and (c) $n=3.2\times 10^{13}\,$cm$^{-3}$. The simulations are performed with 1500~atoms and normalized to the measured distribution. The agreement is excellent for both diffused (left) and unperturbed (right) fractions with  an overlap larger than $90\%$. The unperturbed fraction is mostly composed of tracer atoms that have not suffered any collision (green, simulated).
	}
	\label{fig:density_variation_coll}
\end{figure}

\begin{figure}
	\begin{center}
		\includegraphics{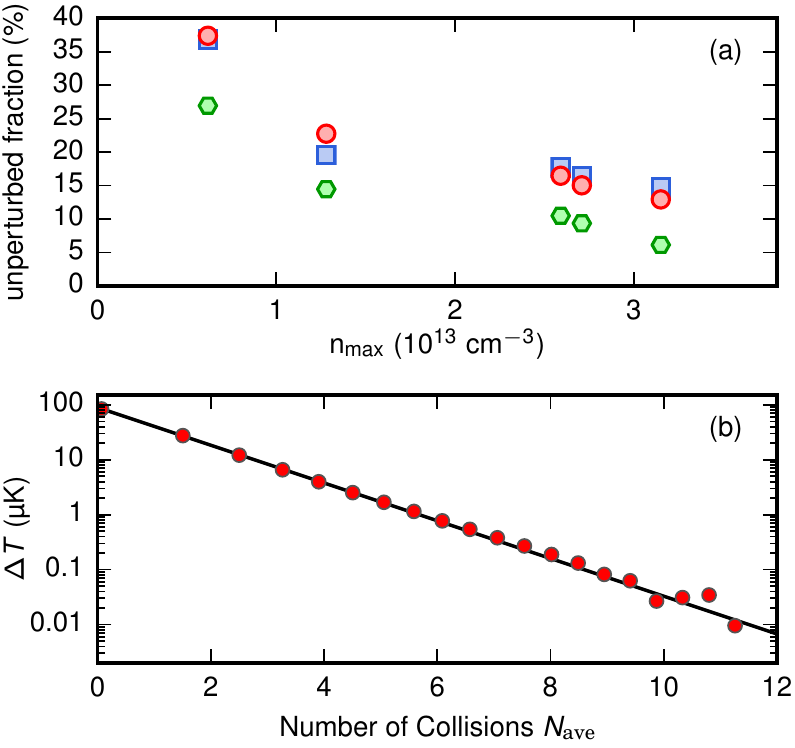}
	\end{center}
	\caption{\textbf{Collision free atoms and thermalization}. (a) Measured (blue squares) and collision simulated (red dots) relative number of atoms in the unperturbed fraction ($z> 0.125\,$mm) for various maximum Rb densities $n_\text{max}$. Good agreement with simulated collision free Cs atoms (green hexagons) is observed. (b) Simulated exponential decay of the difference $\Delta T$ between the Cs kinetic temperature and the Rb temperature, as a function of the average number of collisions $N_\text{ave}$ in a homogeneous cloud. The solid line is an exponential fit to the data with a $1/e$ number of collisions of $1.3$.}
	\label{fig:overlaps}
\end{figure}

In order to quantitatively describe the measured spatial distribution  and the thermalization process, we first employ discrete hard-sphere collision simulations \cite{sup}. We numerically solve Newton's equation of motion for individual Cs tracer atoms elastically colliding  with the thermal gas of Rb atoms in three dimensions. Spatial distributions as shown in Fig.~2 were obtained by projecting the atomic positions onto the $z$-axis. A tagged Cs atom initially starts with an effective Gaussian spatial distribution at $z=\SI{-0.27}{\milli\meter}$ with width $\sigma_x=\sigma_y= \SI{6.8}{\micro\meter}$ and $\sigma_z =\SI{58}{\micro\meter}$ at time $t=\SI{7}{\milli\second}$, and a Maxwell-Boltzmann velocity distribution with temperature $\SI{1.4}{\micro\kelvin}$. These parameters were extracted from an independent time-resolved reference measurement in the absence of the cloud. The properties of the Rb cloud were also obtained by independent measurements, so that there are no free parameters in the simulations. At each time step $dt$, a collision between a tracer atom and a Rb atom is described  by a local collision probability, $P\ti{coll} = 1 - \exp(-\collrate  dt)$, where $\collrate$ is the collision rate. For nonthermal test particles with velocity $v$, the collision rate is given by  (Eq.~5.4,5 in Ref.~\cite{cha70})
\begin{equation}
\collrate(\vcs) = \densrb \frac{\crosssec}{\pi} \sqrt{\frac{2\pi k\tempcloud}{\mrb}}\left[e^{-x^2}+\left(2x+\frac{1}{x}\right)\frac{\sqrt{\pi}}{2}\erf(x)\right],
\label{eq:v_dep_collision_rate}
\end{equation}
where $\sigma$ is the scattering cross section, $x^2 =  {\mrb \vcs^2}/(2k\tempcloud)$  and $\erf(x)$ the error function. We analogously describe three-body losses by the probability $P_\text{loss}= 1- \exp(-dt/\tau)$, with the local lifetime $\tau$ \cite{sup}. Since the density $n$ changes spatially, the expressions above are evaluated at the current position of a tracer.

The results of the discrete collision simulations are compared to experimental tracer distributions in Figs.~2(a)-(c). We obtain excellent agreement between measured (blue) and simulated (red) spatial distributions, both for the unperturbed (right) and the diffused fraction (left), with an overlap, $(\int N\ti{exp} N\ti{sim}dz) /(\int N\ti{exp}^2dz\int N\ti{sim}^2dz)^\frac{1}{2}$, between the two distributions of more than $90\%$. 
We attribute the slight discrepancies to measurement uncertainties of the independently determined parameters entering the simulations \cite{sup}. Figure 3(a) depicts the relative number of atoms in  the unperturbed fraction (defined as those atoms with position $z>0.125$ mm, marked by  the vertical dashed line), for various values of the maximum Rb density $n_\text{max}$.  The simulations show that the unperturbed fraction is mostly composed of tracer atoms that do not undergo any collision (green, simulated). The small difference between the unperturbed and non-collided fractions is due to the sharp cut at $z=\SI{0.125}{\milli\meter}$. In addition,
we find that the computed temperature difference $\Delta T$ between the kinetic temperature of the tagged Cs atoms and the cloud temperature decays exponentially with the number of average collisions $N_\text{ave}$ with a $1/e$ decay constant of 1.3 collisions (Fig.~3 (b)).
This indicates thermalization after a few collisions. We estimate, for example, a relative temperature difference of $\Delta T/T \simeq 2 \%$ after $4 \times 1.3 = 5.2$ collisions.

\begin{figure} [t]
	\begin{center}
		\includegraphics{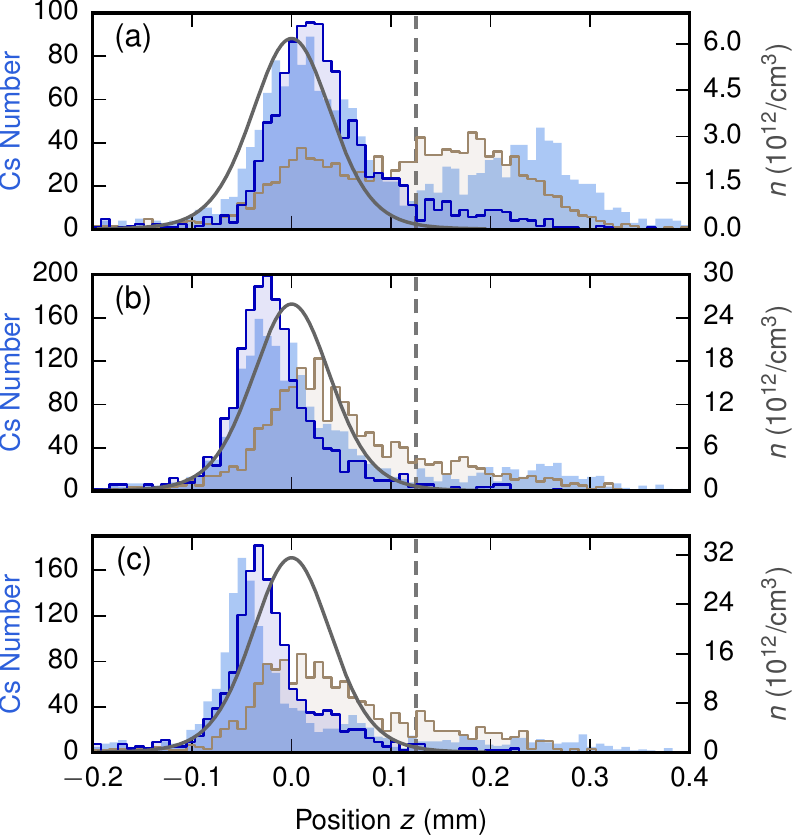}
	\end{center}
	\caption{\textbf{Spatial distribution of Cs atoms - Langevin description}. Measured (blue shaded) and the Langevin simulated spatial distribution (blue line for velocity-dependent friction, brown line for velocity-independent friction \cite{sup}) for the same data as in Fig.~\ref{fig:density_variation_coll}. The simulations are performed with 1500~atoms and normalized to the diffused fraction. Excellent agreement is found for the thermalized diffused fraction (left) between measured data and velocity dependent friction Langevin description, with overlap larger than $90\%$. By contrast, the velocity-independent description yields only poor agreement with measured data with overlaps as low as \SI{60}{\percent}. In both cases, the Langevin equation (2) does not capture the unperturbed fraction (right), as it assumes that all tracer atoms experience damping and noise.}
	\label{fig:density_variation_langevin}
\end{figure}

We next test the validity of the stochastic Langevin approach by simulating  single particle  Langevin trajectories \cite{cof96}. For discrete time steps $dt$, the Langevin equation for the velocity $\vec v$ of a tagged particle takes the form \cite{gil12} 
\begin{equation}
\label{eq:langevinequation}
\vec v(t+dt)=\vec v(t) + \left({\vec F\ti{drag}} + {\vec F\ti{rand}} - {\vec\nabla U}\right) \frac{dt}{m}
\end{equation}
Here,  $\vec F\ti{drag}=-\gamma \vec v$ is a drag force with  friction coefficient $\gamma$, $\vec F\ti{rand}$ a fluctuating force and $-\vec\nabla U$ the confining force including gravity. While  for large, heavy test particles the friction coefficient $\gamma$ is velocity independent, it acquires an explicit  velocity dependence for small, light particles \cite{fer07,fer14}. It was theoretically shown that \cite{fer07,fer14}
\begin{equation}
\gamma(v) = \densrb \crosssec \frac{8}{15} \sqrt{\frac{2\mrb}{\pi \kB T}}\frac{\mcs \left(\mrb v^{2}/2 + 5 \kB T \right)}{\left(\mcs + \mrb\right)} \label{3}
\end{equation}
to order $(M/m)^{3/2}$ and for values of $\mcs \vcs^2/(2\kB\tempcloud)$ not much larger than unity. To our knowledge, the predictions of the Langevin equation with speed-dependent damping coefficient \eqref{3} have  never been verified experimentally so far. We also note that owing to the detailed-balance condition, the nonlinear friction \eqref{3} implies a multiplicative, velocity-dependent fluctuating force \cite{dub09,sup}.

The results of the Langevin simulations are presented in Fig.~4 \cite{sup}. We have again accounted for three-body losses by removing tracer atoms at each time step with the probability $P_\text{loss}$. As before, the cloud density $n$, and thus the friction coefficient $\gamma$, were evaluated at the position of the tagged  atoms. We observe that the diffused fraction of Cs atoms (left) is well described by the Langevin simulations (blue), without free parameters, for all values of the Rb density. However, as expected, the unperturbed fraction (right) is not captured by the Langevin approach which assumes that all test particles experience damping and fluctuations. The overlap between the simulated and measured spatial distributions for the diffused fraction is  larger  than $90 \%$. We once more ascribe the small deviation to the experimental uncertainties of the reference measurements. By contrast, Langevin simulations with the usual velocity-independent friction coefficient (brown) yield overlaps as low as \SI{60}{\percent} \cite{sup}.

The following physical picture emerges from our experimental and theoretical investigations. 
Our ability to accurately identify the effect of a single collision allows  the distinction of two different types of dynamics: Thus, Cs atoms that do not experience any collision pass ballistically through the  dilute Rb cloud (unperturbed fraction). 
On the other hand, a single collision event  dissipates on average more than half of the initial nonthermal kinetic energy and suffices to trap a tagged Cs atom inside the cloud, leading to additional collisions (diffused fraction). 
After few more collisions, a tracer atom will be thermalized with the Rb cloud. Its dynamics is correctly described, over a wide range of Knudsen numbers, by a generalized Langevin equation with a velocity-dependent friction coefficient. The origin of this unfamiliar speed dependence may be  understood by noting that  
 the velocity $v$ of a heavy Brownian particle ($m\gg M$) is much smaller than that of   the gas atoms. As a result, both the collision rate and the friction coefficient are independent of~$v$ in this limit. By contrast, for an atomic tracer with a mass $m\simeq M$, both velocities are of the same order. Here, an explicit velocity dependence of the collision rate and of the friction coefficient can no longer be neglected.
However, as we have shown,  in both cases,  a highly successful continuous Langevin description over a coarse-grained timescale larger than the mean free time between impacts is possible.

We thank Axel Pelster and James Anglin for helpful discussions.
This work was partially funded by the ERC Starting Grant Nr. 278208, the Collaborative Project TherMiQ (Grant Agreement 618074) and the  SFB/TRR49. T. L. acknowledges funding by Carl Zeiss Stiftung, D.M. is a recipient of a DFG-fellowship through the Excellence Initiative by the Graduate School Materials Science in Mainz (GSC 266), F.S. acknowledges funding by the Studienstiftung des deutschen Volkes.
	
\section{Supplementary Material}

\subsection{Knudsen number}
The Knudsen number is defined as $K_n = \lambda/L$, where $\lambda$ is the density-dependent mean-free path within the gas and $L$ is the typical length scale associated with the diffusing particle \cite{bir94}.
For the Cs atoms, the mean free path is given by $\lambda=1/(\sqrt{2}n\sigma)$, where $\sigma = 4 \pi a\ti{Rb,Cs}^2$ is the interspecies scattering cross-section and $a\ti{Rb,Cs}=645\,a_B$~\cite{Takekoshi2012} the $s$-wave scattering length between two Rb atoms ($a_B$ is the Bohr radius). On the other hand, in the absence of the Rb cloud, the dynamics is dominated by collisions with the trapping potential in the radial directions $(x,y)$. We thus identify $L=\sqrt{kT/(M \omega_r^2)}$, the radial $1/e$ width of the thermal Rb cloud with $\omega_r$ the radial angular trap frequency.

\subsection{Negligible quantum effects}
Despite using ultracold temperatures in the  micro Kelvin range, the dynamics  of the atoms in our experiments  can be described classically. 
Quantum effects are  in general relevant when  the condition $ \lambda\ti{th}^3/n \le 1$, with $\lambda\ti{th}$ the thermal de Broglie wavelength, is satified \cite{pet02}.
For the highest Rb density measured, we have instead $ \lambda\ti{th,Rb}^3/n = 16$ for Rb and $ \lambda\ti{th,Cs}^3/n = 30$ for Cs. In addition, the mean-field contribution of the cold Rb cloud to the effective potential on Cs \cite{pet02},
\begin{equation}
V\ti{mf} = n g\ti{RbCs} = n \frac{4 \pi \hbar^2 a\ti{RbCs}}{\mu},
\end{equation}
with $\mu$ the reduced mass, can be used as an order-of-magnitude estimate for the highest Rb density. We have $V\ti{mf}\simeq k\times\SI{0.1}{\micro\kelvin}$, a value which is negligible compared to the potential depth of $k\times\SI{270}{\micro\kelvin}$ for Cs and the bath's thermal energy scale $kT = k\times \SI{2}{\micro\kelvin}$. Furthermore, the dynamics of the Cs atoms in the cloud are not dominated by the trapping potential, but by the Rb density distribution which experiences a six times lower mean field shift due to the lower Rb-Rb scattering length \cite{vanKempen2002}.

\subsection{Three-body loss rate}
\begin{figure}
	\begin{center}
		\includegraphics{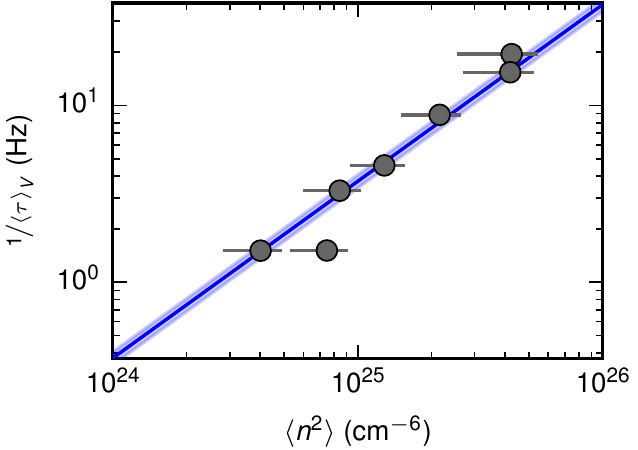}
	\end{center}
	\caption{Measurement of the three-body loss coefficient $L_3$ for Cs in $\ket{F=3}$ and Rb in $\ket{F=1, m_F=0}$
		 Circles: Measured lifetimes.
		Blue solid line: Fit of Eq.~\eqref{eq:lossrate} to the data yielding $L_3 = \SI{37\pm 3}{10^{-26} \centi \meter^6 \second^{-1}}$. The error marks the $1\sigma$ confidence interval originating from statistical uncertainties and is visualized by the blue shaded area.}
	\label{fig:lossrate}
\end{figure}

The lifetime of Cs atoms in our experiment is limited by three-body losses. This influences the evolution of the Cs atom's spatial distribution. Therefore, we have measured the inverse three-body loss rate as a function of Rb density and account for three-body loss in our simulations. The spatially averaged lifetime is~\cite{pet02}, 
\begin{equation}
\frac{1}{\braket{\tau}_V} = L_3 \braket{n\cs \densrb^2}_V
\label{eq:lossrate}
\end{equation}
where $L_3$ is the three-body loss coefficient and $\braket{\cdot}_V$ denotes the spatial average over the cloud's volume $V$.
The local lifetime of a Cs atom is proportional to $\densrb^2$, so Cs loss predominantly happens at the center of the Rb cloud.
 To   determine the coefficient $L_3$, we measure $\braket{\tau}_V$  for various Rb densities by inserting Cs atoms into the  cold cloud and recording the spatial distribution at various interaction times.
Fitting Eq.~(\ref{eq:lossrate}) to the experimental data, as shown in Fig.~\ref{fig:lossrate}, yields $L_3=\SI{37\pm 3}{10^{-26} \centi\meter^6 \second^{-1}}$.
Here, Cs is prepared in $\ket{F=3}$ and Rb in $\ket{F=1, m_F=0}$ and the loss coefficient is roughly one order of magnitude larger than for Cs in the $\ket{F=3,m_F=3}$ and Rb in the $\ket{F=1,m_F=1}$ state \cite{Spethmann2012}.
Using the local lifetime $\tau (n)$ as a function of density, allows accounting for three-body recombination in our simulations.

\subsection{Discrete collision simulations}
In this section, we present details of the hard-sphere collision simulations. We describe Cs tracer atoms as point-like masses that evolve in the absence of a collision according to Newton's equation of motion, $\vec v(t+dt)=\vec v(t) - {\vec\nabla U} dt/m$, during each time step $dt$, where $U$ is the confining potential including gravity. We model the Rb cloud as a thermalized classical gas with time-independent density $n$ and temperature $T$. This assumption is justified in view of the extreme number ratio between Cs and Rb atoms. 
At each time step, the local collision probability $P\ti{coll} = 1 - \exp(-\collrate dt)$ with the position and velocity dependent collision rate $\collrate$ is computed. Comparing $P\ti{coll}$ to a uniform random number determines if the collision occurs.
For each collision event, a Rb collision partner is randomly generated {and the Cs atom is assigned a new velocity $\vcs$ by solving the two-body collision in the center-of-mass system under the assumption of isotropic $s$-wave scattering. This assumption is compatible with our hard-sphere model and well justified for the kinetic energies involved. The Rb atom does not play any further role in the simulation as a new, statistically representative Rb atom is drawn for each collision. Such a procedure is justified because the gas is dilute and correlations due to multiple scatterings with the same Rb atom can be neglected.  We account for three-body losses with the probability $P_\text{loss}= 1- \exp(-dt/\tau)$, with the local lifetime $\tau$.  In view of the initial nonthermal kinetic energy of the tagged atoms, the velocity $\mathbf{\vrb}$ of the Rb collision particle before the collision should be determined with care. Drawing the velocity of the Rb atom from a Maxwell-Boltzmann distribution is flawed, as the speed-dependence of $\collrate$ would depopulate high-velocity classes of $\vcs$ due to more frequent scattering and prevent thermalization. This is the same reason, why the fluctuating force in the Langevin equation becomes speed dependent. The Rb velocity should instead be drawn from a modified  (not normalized) Maxwell-Boltzmann distribution given by (see Ref.~\cite{cha70}, Chapter 5.4),
	\begin{eqnarray}
	P_{\vrb}(\vcs) &=& \pi \int_{0}^{\pi} g(\vrb, \vcs, \theta) \sin \theta d\theta\int_{0}^{2\pi}d\phi \frac{\crosssec}{\pi} f_\mrb \vrb^2\\
	P_{\theta}(\vcs,\vrb) &=& g(\vrb, \vcs, \theta) \sin \theta\\
	P_{\phi} &=& 1.
	\end{eqnarray}
	where $\vrb$, $\theta$ and $\phi$ are polar coordinates of the Rb speed vector about $\mathbf{\vcs}$ as an axis with probability distributions $P_{\vrb}$, $P_{\theta}$, $P_{\phi}$; $g$ is the relative velocity of the two collision partners and $f_\mrb$ the usual Maxwell-Boltzmann-distribution
	\begin{equation}
	f_\mrb = \densrb \left(\frac{\mrb}{2\pi k \tempcloud}\right)^\frac{3}{2} \exp\left(\frac{-\mrb \vrb^2}{2k\tempcloud}\right).
	\end{equation}
\begin{figure}[t]
	\begin{center}
		\includegraphics{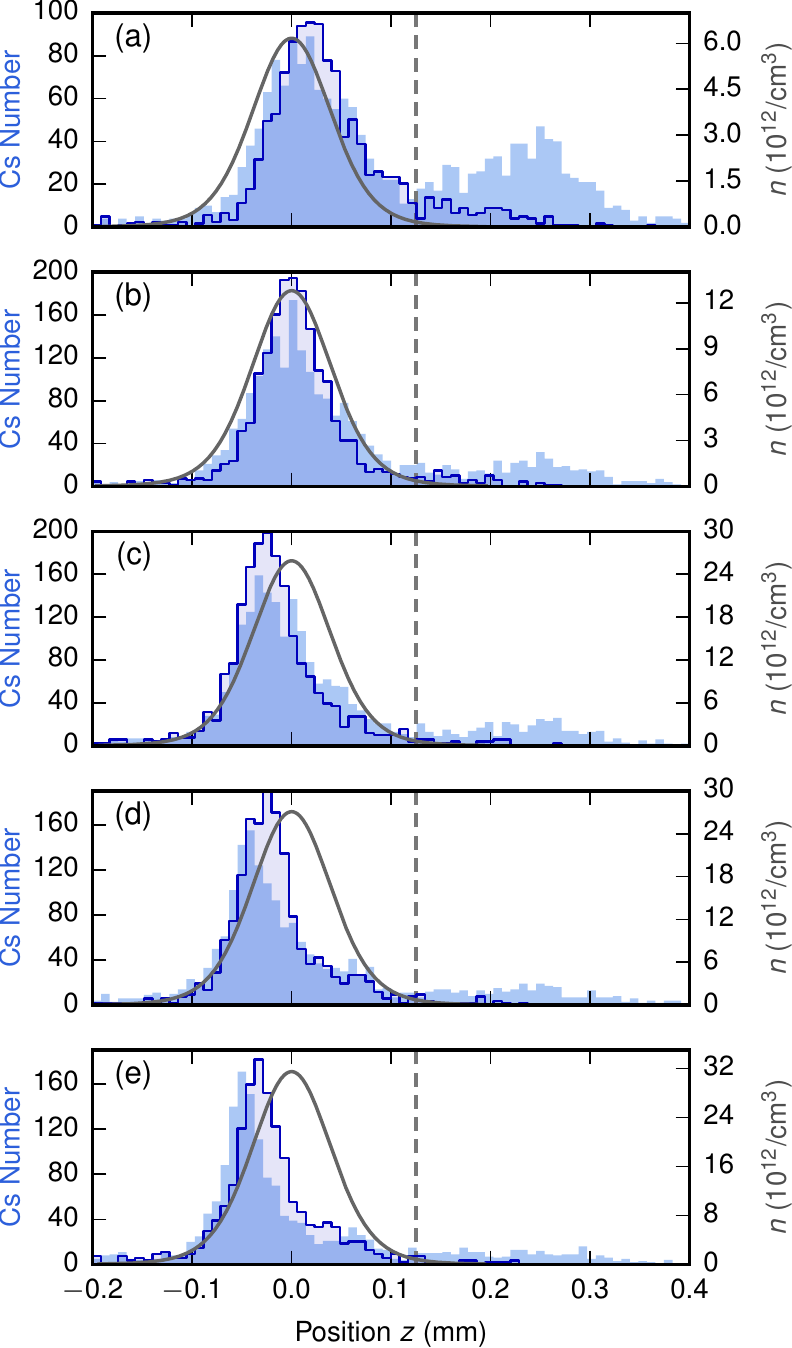}
	\end{center}
	\caption{Measured Cs distributions (blue histograms, borderless) for various peak Rb densities at $t=\SI{25}{\milli\second}$ with Langevin simulated distributions (blue histogram, solid border), normalized to the diffused fraction.}
	\label{fig:density_variation_langevin_all}
\end{figure}
\begin{figure}[t]
	\begin{center}
		\includegraphics{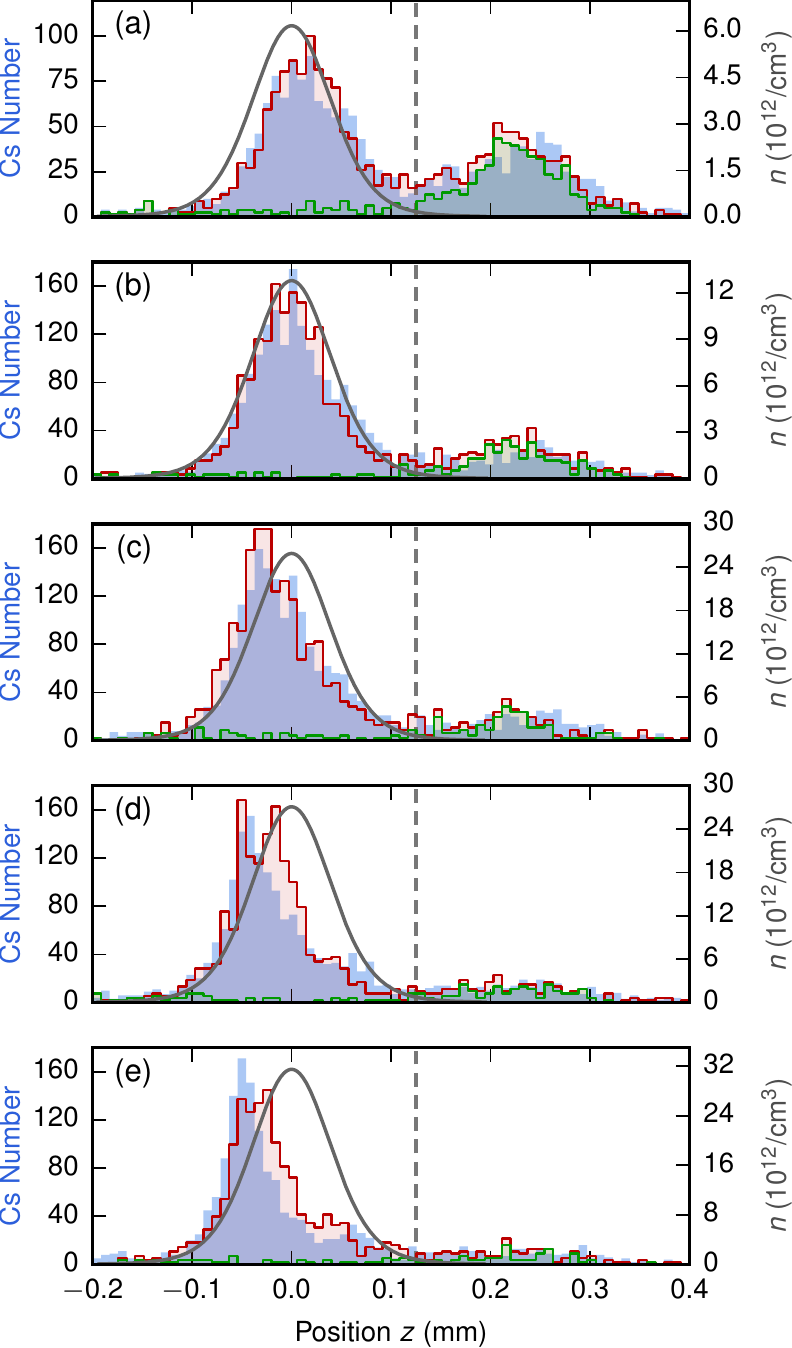}
	\end{center}
	\caption{Measured Cs distributions (blue histograms, borderless) for various peak Rb densities at $t=\SI{25}{\milli\second}$ with collision simulated distributions (red histogram, solid border), normalized to the diffused fraction.
	In green (solid border), the fraction of atoms that have not underwent any collision are shown.
	}
	\label{fig:density_variation_coll_all}
\end{figure}
\subsection{Langevin simulations}
We numerically solve the  Langevin equation using a 3rd-order Runge-Kutta integrator with a typical time step of $dt=\SI{2}{\micro\second}$ and 1500 trajectories, corresponding to the number of observed Cs atoms. The fluctuation-dissipation relation for an equilibrium bath implies a direct  connection between damping and noise terms.  For the velocity-dependent friction coefficient,\begin{equation}
\gamma(v) = \densrb \crosssec \frac{8}{15} \sqrt{\frac{2\mrb}{\pi \kB T}}\frac{\mcs \left(\mrb v^{2}/2 + 5 \kB T \right)}{\left(\mcs + \mrb\right)} 
\end{equation}
the detailed-balance condition and the Gibbs form of the thermal equilibrium condition lead to the speed-dependent random force $\vec F\ti{rand} = \psi(v) \sqrt{{2}/{dt}}\vec N$, where $\vec N$ is a normally distributed random force with zero mean and unit variance, and \cite{dub09}
\begin{equation}
\begin{split}
\psi^2(v) = &2\mcs e^{\frac{\mcs v^2}{kT}} \int_{v}^{\infty} -\gamma(u) u e^{-\frac{\mcs u^2}{kT}} du\\
=& 2^{\frac{5}{2}}  \densrb \crosssec \sqrt{\frac{\mrb kT}{\pi}} \frac{\mrb \mcs v^{2} + \left(\mrb + 10 \mcs \right) k T }{15(\mrb + \mcs)}.
\end{split} \label{11}
\end{equation}
Equation \eqref{11} corresponds to multiplicative noise \cite{san82}.
On the other hand, for the usual Langevin equation the friction coefficient is velocity independent and reads
\cite{fer07, fer00},
\begin{equation}
\gamma =  n \sigma \frac{8}{3} \sqrt{\frac{2MkT}{\pi}} \frac{m }{\left(m + M\right)}
\end{equation}
The random force is accordingly characterized by
\begin{equation}
\psi^2 = m \gamma k T \label{12}
\end{equation}
Equation \eqref{12} corresponds to additive noise \cite{san82}.

\begin{figure}
	\begin{center}
		\includegraphics{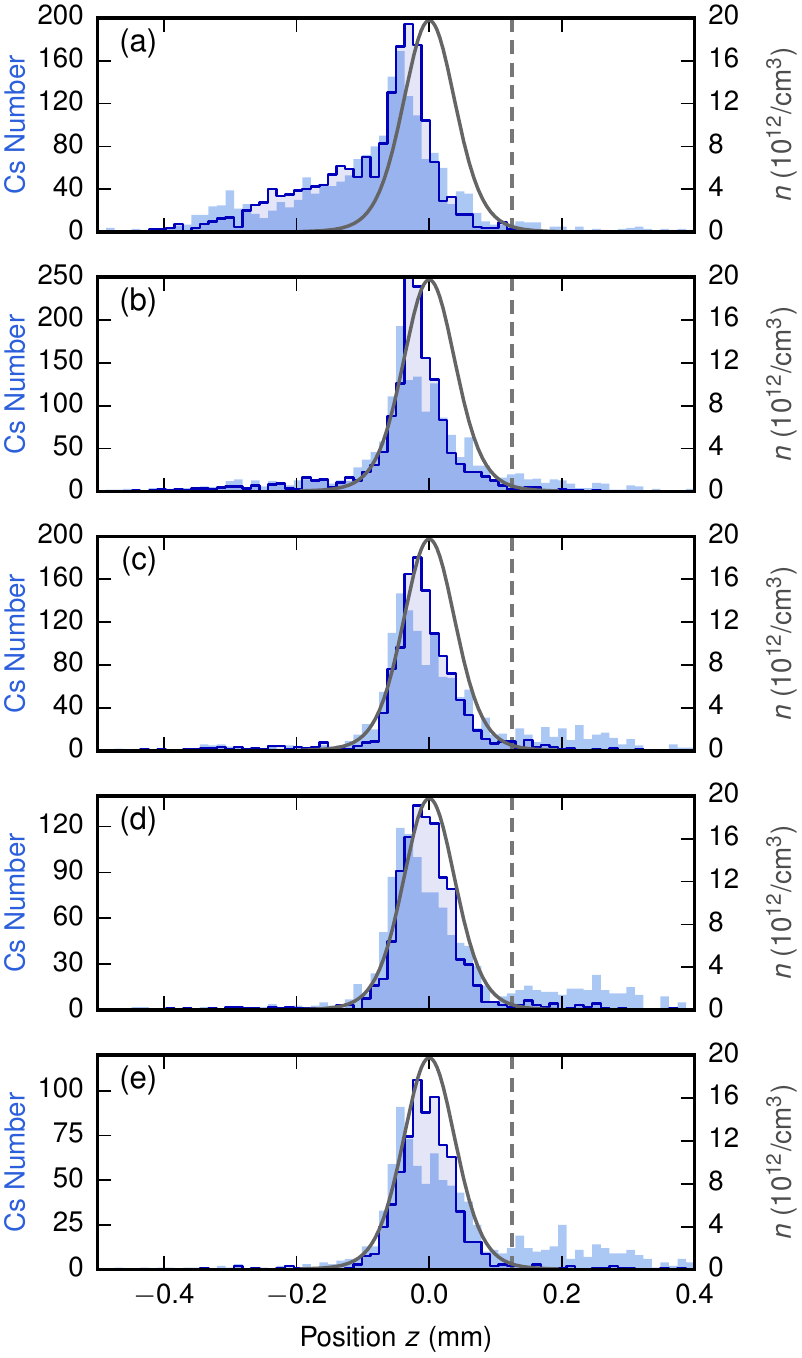}
	\end{center}
	\caption{Measured Cs distributions (blue histograms, borderless) for a fixed Rb density and growing interaction time $t=17, 21, 25, 29, 33 \si{\milli\second}$ from (a)-(e) with Langevin simulated distributions (blue histogram, solid border), normalized to the diffused fraction.}
	\label{fig:time_variation_langevin}
\end{figure}
\begin{figure}
	\begin{center}
		\includegraphics{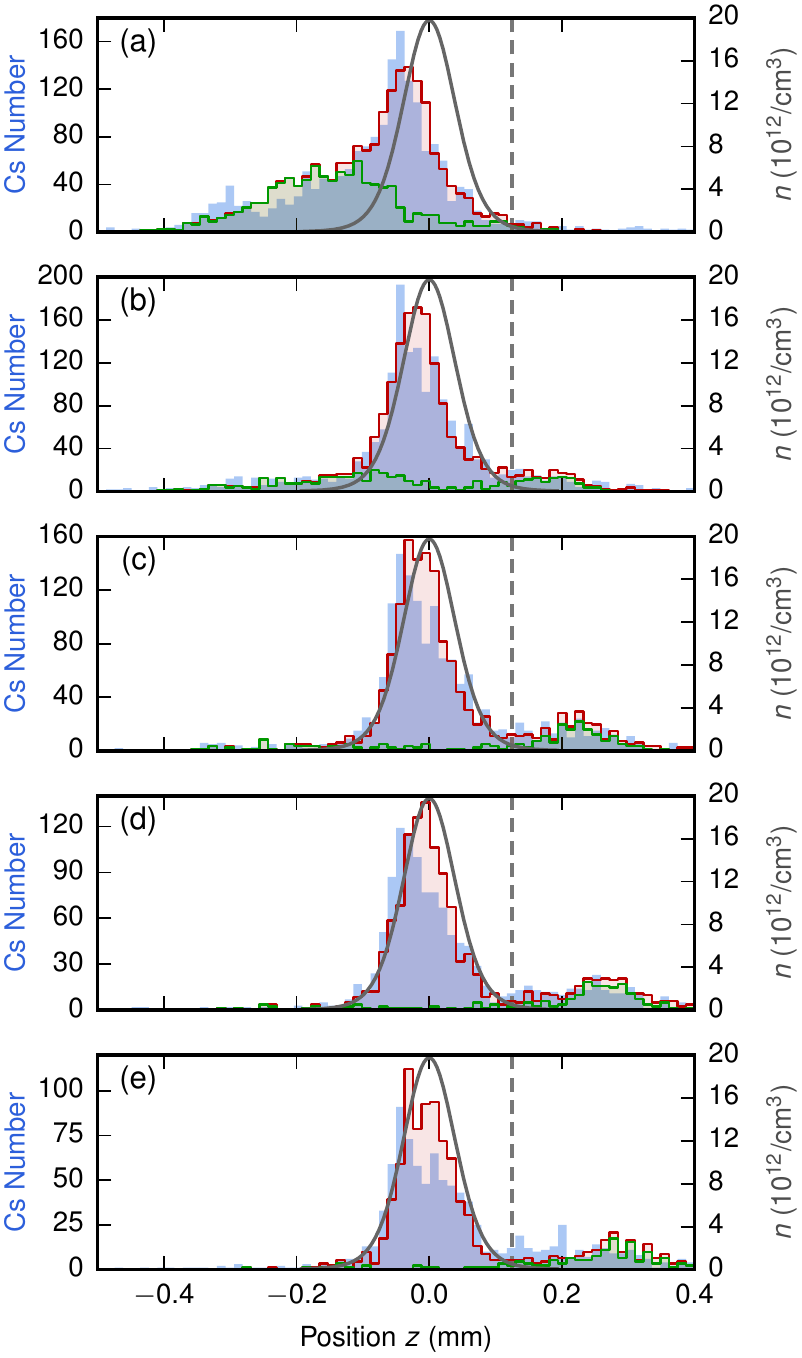}
	\end{center}
	\caption{Measured Cs distributions (blue histograms, borderless) for a fixed Rb density and growing interaction time $t=17, 21, 25, 29, 33 \si{\milli\second}$ from (a)-(e) with collision simulated distributions (red histogram, solid border), normalized to the diffused fraction.
	In green (solid border), the fraction of atoms that have not underwent any collision are shown.
	}
	\label{fig:time_variation_coll}
\end{figure}
\subsection{Simulation parameters}

	Our numerical simulations  are conducted without free parameters. The simulation parameters are obtained from independent measurements. The trapping potential consists of two Gaussian beams at a wavelength of $\lambda=\SI{1064}{\nano\meter}$ and the potential is precisely known from beam profile measurements, consistent with trap-frequency measurements with Rb and measured single Cs atom motion in absence of Rb.
	The vertical beam along the $x$ axis has a power of $\SI{4}{\watt}$. Its focus is located at $x=\SI{4}{\centi\meter}$ and its beam waist is $\SI{0.16}{\milli\meter}$, where the intersection of the two beams defines the coordinate system's origin. The horizontal beam along $z$ is elliptical with beam waists of $w_x= \SI{21}{\micro\meter}$ and $w_y=\SI{22}{\micro\meter}$, a power of $\SI{680}{\milli\watt}$, and its focus at $z=0$. The beams produce a potential per intensity $I$ of $U/I=\SI{-3.60e-36}{\joule \meter^2 \per \watt}$ for Cs and $U\rb/I=\SI{-2.10e-36}{\joule \meter^2 \per \watt}$ for Rb.
	
	The thermal distribution of Rb in the trap is calculated according to Boltzmann's law 
	\begin{equation}
		\densrb(\vec{r}) = \numrb e^{\frac{-U\rb(\vec{r})}{k\tempcloud}}\left(\int{ e^{\frac{-U\rb(\vec{r})}{k\tempcloud}}}\right)^{-1}
	\end{equation}
	\begin{figure}[H]
		\begin{center}
			\includegraphics{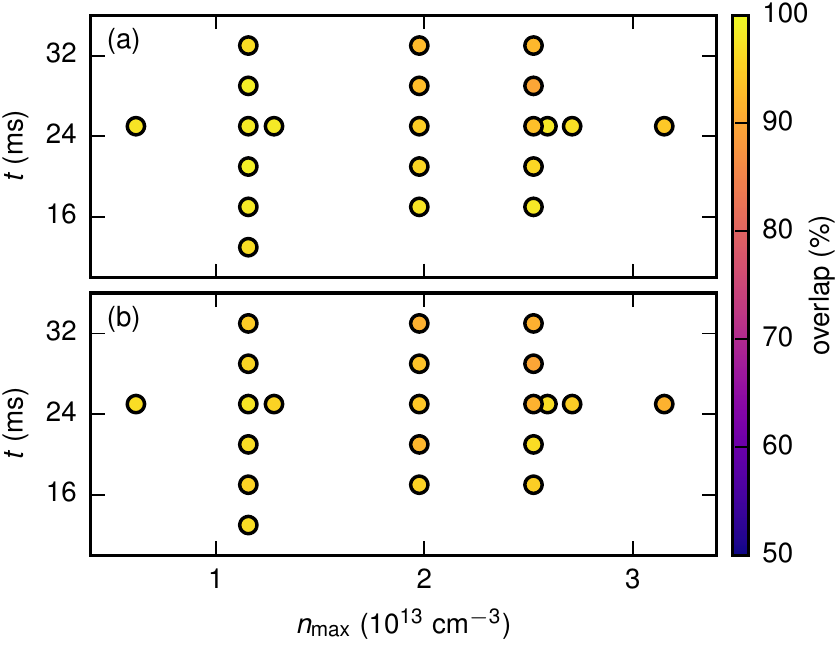}
		\end{center}
		\caption{Overlaps $(\int N\ti{exp} N\ti{sim}dz) /(\int N\ti{exp}^2dz\int N\ti{sim}^2dz)^\frac{1}{2}$ for all tracer diffusion measurements at peak densities $n\ti{max}$ and interaction times $t$.
			Overlaps for the hard-sphere collision simulation are shown in (a), overlaps for the Langevin simulation with velocity-dependent friction, computed for the diffused fraction, are shown in (b). All overlaps for both models are $\ge \SI{90}{\percent}$, whereas the Langevin simulation with velocity-independent friction shown in Fig.~4 of the main article yields overlaps down to $\SI{60}{\percent}$.	}
		\label{fig:overlaps_2D}
	\end{figure}
	\noindent while $\tempcloud$ and the Rb atom number $N$ are determined with time-of-flight thermometry and absorption imaging~\cite{Ket98}.
	It is important to note that a harmonic approximation of the trapping potential does not hold. We  consider instead} the exact Gaussian beam potential throughout the simulation.
	The interspecies scattering length $a\ti{Rb,Cs}=645\,a_B$~\cite{Takekoshi2012} for Rb in $\ket{F=1,m_F=1}$ and Cs in $\ket{F=3,m_F=3}$ is used as an approximation for the different hyperfine states employed in this experiment.

\subsection{Detailed density distribution measurements}
The full measurement of the Cs spatial distribution when the Rb density is varied at a fixed interaction time corresponding to Figs.~\ref{fig:density_variation_coll}-\ref{fig:density_variation_langevin} is shown in Figs.~\ref{fig:density_variation_langevin_all}-\ref{fig:density_variation_coll_all}, together with Langevin and  discrete hard-sphere collision simulations.
Spatial distributions for various interaction times at a fixed Rb density, corresponding to Fig.~\ref{fig:Sketch}, are shown in Figs.~\ref{fig:time_variation_langevin}-\ref{fig:time_variation_coll}. A comparison of the overlaps for a full, consistent measurement of different densities and interaction times is depicted in Fig.~\ref{fig:overlaps_2D}, where the overlap of measured data with both the collision simulation (Fig.~\ref{fig:overlaps_2D}(a)) and  the Langevin model with velocity-dependent friction coefficient (Fig.~\ref{fig:overlaps_2D}(b)) yields values $\ge\SI{90}{\percent}$. By contrast, a Langevin model with velocity-independent friction coefficient yields typical overlaps of $\approx60\%$. We emphasize that the simulations have been performed with a single, consistent set of independently measured parameters for all data, without free parameter.

\end{document}